\documentclass{article}

\usepackage[final]{neurips_2025}

\usepackage[utf8]{inputenc} 
\usepackage[T1]{fontenc}    
\usepackage{hyperref}       
\usepackage{url}            
\usepackage{booktabs}       
\usepackage{amsfonts}       
\usepackage{nicefrac}       
\usepackage{microtype}      
\usepackage{xcolor}         
\usepackage{graphicx}

\title{Reciprocity as the Foundational Substrate of Society: How Reciprocal Dynamics Scale into Social Systems}

\author{%
  Egil Diau \\
  Department of Computer Science\\
  Nation Taiwan University\\
  Taiwan, Taipei \\
  \texttt{egil158@gmail.com} \\
}

\begin{document}

\maketitle

\begin{abstract}
Prevailing accounts in both multi-agent AI and the social sciences explain social structure through top-down abstractions—such as “institutions,” “norms,” or “trust”—yet lack simulateable models of how such structures emerge from individual behavior. Ethnographic and archaeological evidence suggests that reciprocity served as the foundational mechanism of early human societies, enabling economic circulation, social cohesion, and interpersonal obligation long before the rise of formal institutions. Modern financial systems such as credit and currency can likewise be viewed as scalable extensions of reciprocity, formalizing exchange across time and anonymity. Building on this insight, we argue that reciprocity is not merely a local or primitive exchange heuristic, but the scalable substrate from which large-scale social structures can emerge. We propose a three-stage framework to model this emergence: reciprocal dynamics at the individual level, norm stabilization through shared expectations, and the construction of durable institutional patterns. This approach offers a cognitively minimal, behaviorally grounded foundation for simulating how large-scale social systems can emerge from decentralized reciprocal interaction.
\end{abstract}

\begin{figure}[htbp]
  \centering
  \includegraphics[width=0.8\textwidth]{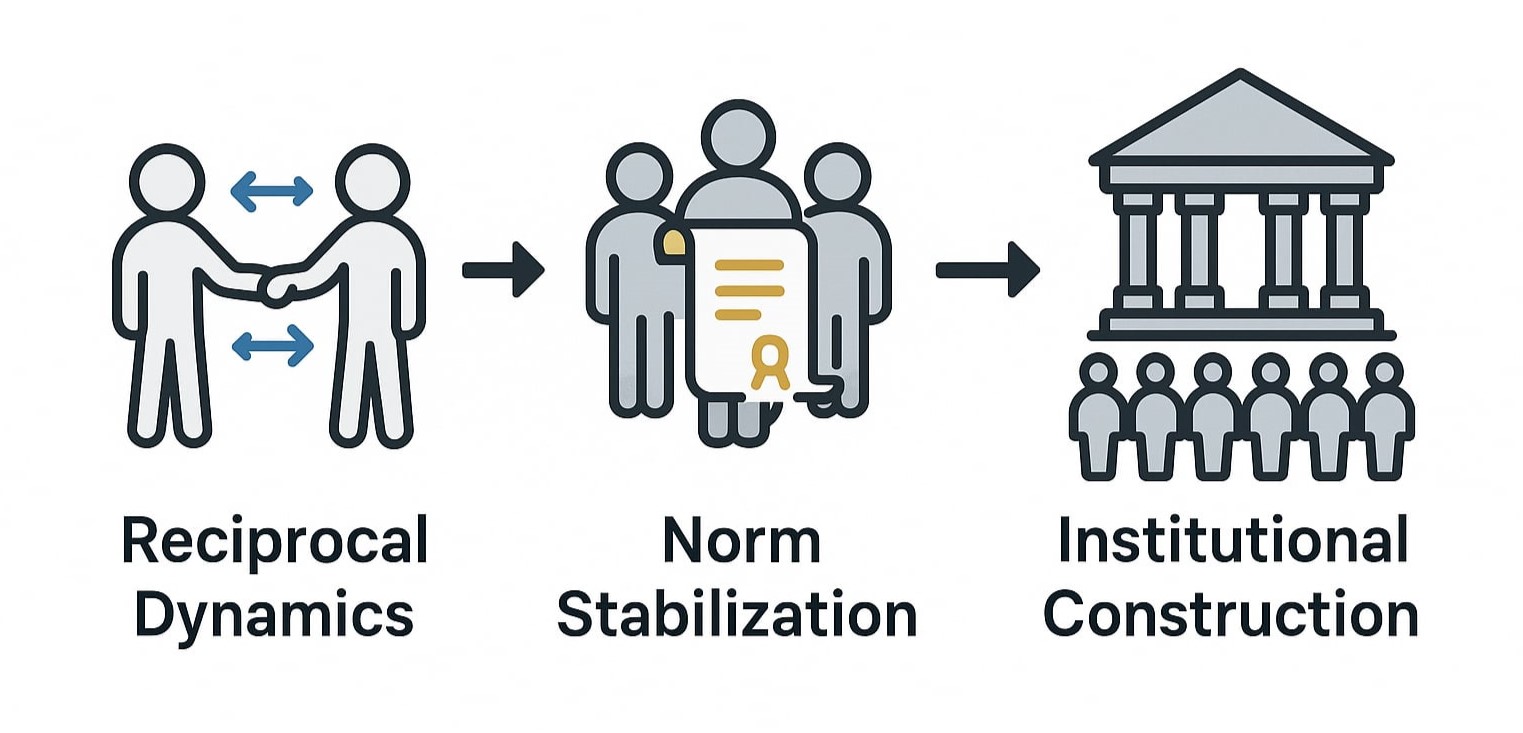}
  \caption{We propose a three-stage framework in which scalable institutions emerge from decentralized reciprocal interaction—if systems like early exchange and modern finance are understood as extensions of reciprocity.}
  \label{fig:main}
\end{figure}

\section{Introduction}

In multi-agent AI, the formation of durable social structures remains poorly understood. Although cooperation and communication have been extensively modeled, existing approaches typically rely on top-down assumptions—such as shared goals, norms, or symbolic agreement—without explaining how these emerge from individual interactions.

This limitation reflects a longstanding gap in economics and sociology, where institutions, norms, and markets are often invoked as explanatory primitives rather than derived from agent-level behavior. Their origins and operational dynamics remain underspecified, constraining both theoretical clarity and simulation potential.

Across early human societies, \textit{reciprocity} provided the behavioral foundation for cooperation, resource flow, and social cohesion \cite{sahlins2013stone, mauss2024gift, malinowski2013argonauts}. Long before the emergence of formal institutions, group life was structured through dynamic, memory-based reciprocal exchange. 

We further observe that reciprocity is not confined to small-scale or informal settings. Credit, currency, and capital transfer can be viewed as scalable forms of reciprocity, enabling cooperation across time, space, and anonymity. 

Yet despite its presence across both early and modern systems, we lack a formal understanding of how reciprocity scales into large-scale social structures. Predominant theories often treat institutions, norms, and markets as pre-given abstractions, without reconstructing how they could emerge from decentralized reciprocal interaction. This leaves a critical explanatory gap: how do individually grounded exchanges give rise to the stable, impersonal coordination seen in large societies?

To address this gap, we propose a three-stage bottom-up framework for the emergence of social structures from minimal cognitive-behavioral substrates. The framework proceeds through:
\begin{enumerate}
    \item \textbf{Reciprocal Dynamics:} Emergence of minimal social units that enable value or obligation to be transferred across agents over time.
    \item \textbf{Norm Stabilization:} Consolidation of these patterns into soft, population-level expectations that regulate behavior.
    \item \textbf{Institutional Construction:} Externalization of stabilized expectations into fixed, environment-level constraints that structure interaction at scale.
\end{enumerate}

Unlike traditional theories that assume the existence of social order, our approach reconstructs social structure as an emergent property of minimally capable agents interacting over time. By grounding emergence in basic cognitive dynamics, we enable the study of how complex moral, cultural, and institutional systems can arise naturally, without presupposing symbolic reasoning or centralized authority.

\paragraph{Our contribution.}
Our contribution is not an engineering system or a descriptive ethnography. It is a theoretical realignment: an attempt to reconstruct the origins of social structures from cognitively minimal, behaviorally plausible substrates. Rather than presupposing norms, institutions, or markets, we seek to simulate their natural emergence through basic agent-level mechanisms. Specifically, we:

\begin{itemize}
    \item Identify reciprocity as a scalable behavioral substrate of social structure, bridging small-scale interaction and large-scale institutional systems;
    \item Introduce a three-stage framework—Reciprocal Dynamics, Norm Stabilization, and Institutional Construction—for modeling the emergence of social structure from minimal agent-level behavior;
    \item Provide a simulateable and cognitively grounded model, suitable for agent-based implementation and computational validation;
    \item Reframe “norms” and “institutions” as emergent macrostates—rather than fixed primitives—arising from distributed reciprocal behavior over time.
\end{itemize}

\paragraph{Ethical Statement}
This work does not rely on evolutionary explanations or biological determinism.
While we draw on behavioral evidence from primates and early human societies, our goal is not to claim innate or adaptive origins of social structures.
Rather, these observations serve as empirical constraints to inform the design of minimal cognitive and behavioral mechanisms sufficient for the emergence of reciprocal norms and institutions.
Our account focuses on behavioral plausibility and simulateable processes, without assuming evolutionary teleology, cultural essentialism, or autonomous social evolution.

\section{Related Work}
\label{gen_inst}

\subsection{Agent-Based Modeling of Social Behaviors}

Recent advances in multi-agent systems have enabled negotiation, planning, and scripted cooperation \citep{park2023generative, li2023camel, leibo2017multi}. However, such systems often lack behavioral grounding: interactions are typically driven by optimization objectives or surface-level heuristics, without modeling the internal cognitive processes—such as memory, expectation formation, or social contingency—that stabilize long-term social behavior.

Attempts to introduce episodic memory or heuristics remain cognitively unprincipled, leading to unstable or hardcoded outcomes \citep{baker2019emergent}. As a result, cooperation typically emerges from centralized engineering—not from decentralized inference or behavioral structure.

This limits the ability to simulate how real-world social systems arise from low-level interactions. Our framework fills this gap by introducing a minimal set of simulateable, behaviorally plausible mechanisms that support the emergence of scalable social organization without presupposing symbolic reasoning or centralized control.

\subsection{Behavioral Foundations of Reciprocity: Insights from Primatology and Anthropology}

Empirical studies in primatology suggest that reciprocal behaviors—such as grooming, food sharing, and alliance formation—are central to the maintenance of long-term social bonds and group stability among primates \cite{de1997chimpanzee}.

Anthropological studies similarly identify reciprocity as the foundational structure of early human societies, underpinning economic exchange, social organization, and interpersonal relationships \cite{sahlins2013stone, mauss2024gift}. Contrary to the enduring myth of barter, early exchange systems were structured around delayed return and embedded obligation rather than direct equivalence.

Building on these findings, we formalize a minimal framework for how reciprocal interactions can give rise to stable norms and scalable institutions under cognitively realistic conditions.

\subsection{Classical Social and Economic Theories: Presupposed Structures}

Traditional theories in sociology and economics often treat institutions, norms, and markets as explanatory primitives—emergent from collective belief, rational design, or functional necessity—without reconstructing how such structures arise from individual-level behavior \cite{durkheim2023rules, weber1978economy, north1990institutions, ostrom1990governing}.

As a result, these models offer post hoc descriptions rather than simulateable accounts of emergence. Our framework addresses this gap by modeling institutions as environment-level macrostates arising from minimal reciprocal interaction under cognitive and behavioral constraints.


\section{Background: From Presupposed Structures to Emergent Reciprocity}
\label{headings}

\subsection{Presupposed Structures in Social and Economic Thought}

Sociological and economic theories often treat norms, markets, and institutions as given structures that shape behavior. Yet few address how such structures emerge from decentralized interaction.

In sociology, Durkheim’s “social facts” \cite{durkheim2023rules} and Parsons’ systems theory \cite{parsons2013social} assume that institutions precede agency. Even interpretive approaches rarely explain how regularities arise from minimal behavior. In economics, neoclassical models presuppose markets and property rights, while institutional theories such as Coase and North \cite{coase2013problem, north1990institutions} treat governance as exogenous—without modeling its behavioral origins.

This leaves a critical gap: how can complex social structures emerge from minimal behavioral mechanisms—without presupposing shared norms, authority, or symbolic agreement?

\subsection{Toward a Bottom-Up Reconstruction of Social Structures}

If early societies were sustained by reciprocal exchange, and modern systems—from credit to currency—function as scalable extensions of that logic, then reciprocity may serve as a fundamental substrate that enables society to operate at scale.

While many social norms arise from biologically grounded predispositions—such as norm aversion, in-group preference, or dominance-seeking—these mechanisms tend to stabilize behavior only within bounded contexts. They often lack the flexibility to support cooperation across roles, time, and social distance. In contrast, reciprocity enables delayed, asymmetric, and partner-contingent interaction, making it uniquely suited to scale with growing social complexity.

\paragraph{Our Framework.} To formalize this perspective, we introduce a three-stage framework for modeling the emergence of social structure from minimal reciprocal interaction. Each stage builds on the last, gradually transforming local exchanges into abstract and scalable forms of coordination:

\begin{enumerate}
    \item \textbf{Reciprocal Dynamics:} Emergence of minimal social units that enable value or obligation to be transferred across agents over time.
    \item \textbf{Norm Stabilization:} Consolidation of these patterns into soft, population-level expectations that regulate behavior.
    \item \textbf{Institutional Construction:} Externalization of stabilized expectations into fixed, environment-level constraints that structure interaction at scale.
\end{enumerate}

In this framing, while norms may originate from various sources, only reciprocity provides the behavioral substrate capable of sustaining scalable, flexible social structures.

\section{Theoretical Framework: From Reciprocal Dynamics to Institutional Construction}

\subsection{Reciprocal Dynamics: Emergence of Agent-Level Reciprocity}

\subsubsection{General Concept and Mechanistic Inference}

Reciprocity is a biologically grounded foundation for social interaction—enabling cooperation, exchange, and coordination. It is sustained by two core motivations: the felt obligation to return past benefits (indebtedness) and the intrinsic satisfaction of helping others. These motivations emerge early in development, appear across cultures, and are widely observed in nonhuman primates.

Their expression likely depends on behavioral capacities that are themselves shaped by the demands of group living—such as the need to distinguish social partners, remember past encounters, and evaluate whether ongoing interactions are beneficial over time. These capacities emerge naturally through repeated social interaction and are common across many group-living species.

\subsubsection{Historical Illustration and Support}

Reciprocal interaction has long been recognized as a foundational feature of human social life—and finds clear parallels in nonhuman primates. Chimpanzees and bonobos, for instance, engage in reciprocal behaviors such as grooming, food sharing, and coalition support \cite{de1997chimpanzee, waal2007chimpanzee}, suggesting that reciprocity is not a product of human culture, but a biologically grounded capacity shared across species.

In early human societies, reciprocity structured the core of exchange, cooperation, and social bonding—well before the emergence of formal institutions. Sahlins \cite{sahlins2013stone} classified reciprocal forms into generalized, balanced, and negative modes based on social distance. Malinowski \cite{malinowski2013argonauts} showed how prestige goods circulated in the Kula ring through enduring reciprocal ties. Mauss \cite{mauss2024gift} formalized gift exchange as a triadic process of giving, receiving, and returning embedded in ongoing relationships.

In short, reciprocity was not a cultural invention, but a behavioral foundation for early human society—enabling durable social coordination without formal rules.

\subsubsection{Simulation Pathways and Implementation}

To implement reciprocal dynamics in LLM-based agents, this design builds on prior work \cite{diau2025cognitivefoundationseconomicexchange} and operationalizes three core behavioral primitives using lightweight memory and interaction logic. Without relying on reward engineering or predefined cooperation rules, agents interact through minimal structures that support the emergence of reciprocal behavior:

\begin{itemize}
    \item \textbf{Partner differentiation:} Agents can distinguish between different social partners or partner types and maintain corresponding interaction histories.
    
    \item \textbf{Reciprocal evaluation heuristics:} For each partner, agents estimate the likelihood of future reciprocity by integrating past positive interactions and contextual signals. This credence modulates the agent’s willingness to initiate or sustain cooperative behavior.
    
    \item \textbf{Behavioral updating:} Agents monitor the balance of cooperative investment over time and adapt their behavior based on perceived asymmetries—reducing cooperation when consistently exploited and increasing it when returns are favorable.
\end{itemize}

This minimal setup enables agents to form stable reciprocal relationships without centralized coordination. Even with sparse structure, agents can self-organize into cooperative networks, providing a foundation for more complex social dynamics in later stages of simulation.

\subsection{Norm Stabilization: Reciprocity as Shared Social Expectations}

\subsubsection{General Concept and Mechanistic Inference}

Repeated reciprocal interactions can lead to the emergence of shared behavioral expectations. As agents adjust their behavior based on prior outcomes and partner responses, certain patterns become statistically predictable and socially reinforced. Over time, these expectations stabilize into norms—implicit rules that guide action even in the absence of immediate return.

Such norms reduce uncertainty, bias decision-making toward cooperative behavior, and extend reciprocal coordination beyond dyadic memory. By anchoring behavior in socially shared expectations, they provide a foundation for more abstract and scalable forms of social organization.

\subsubsection{Historical Illustration and Support}

Cross-cultural examples show how repeated reciprocal interactions can lead to shared expectations about fairness and obligation. In the Trobriand Islands, the Kula ring system involved long-distance, delayed exchanges of prestige items, where participants were expected to return gifts over time and maintain reputational standing \cite{malinowski2013argonauts}. By contrast, the !Kung San of southern Africa share meat immediately and equally after a hunt, with strong informal norms discouraging hoarding or showing off \cite{lee1979kung}.

Both cases demonstrate how reciprocity, whether delayed or immediate, can give rise to norms that regulate exchange, enforce social balance, and support broader cooperation.

\subsubsection{Simulation Pathways and Implementation}
Norms are implemented as dynamic, environment-level macrostates that emerge from decentralized reciprocal interactions. Rather than being explicitly coded or centrally enforced, normative patterns arise as soft statistical regularities based on agent behavior. The implementation can be specified as follows:

\begin{itemize}
    \item \textbf{Normative macrostate:} An environment-level distribution capturing the prevalence of cooperative versus non-cooperative interactions across the population. Agents do not perceive this macrostate directly; rather, they adjust behaviors based on local interaction histories, with the macrostate emerging as an aggregate outcome.
\end{itemize}

As these individual-level adjustments accumulate, local reciprocity patterns begin to converge—giving rise to stable, population-wide expectations that function as emergent norms.

\subsection{Institutional Construction: Money and Rules as Scalable Reciprocity}

\subsubsection{General Concept and Mechanistic Inference}

Some norms, once stabilized through repeated reciprocal interactions, begin to function as shared expectations. When these expectations are externalized—through codified roles, rules, or record-keeping—they become persistent constraints on behavior. This process transforms informal norms into formal institutions, enabling coordinated action at scale across time and social distance.

Modern economic systems exemplify this transformation. Money, in particular, serves as a scalable abstraction of reciprocity—decoupling exchange from direct interaction and individual memory. By acting as a transferable placeholder for obligation, it enables cooperation across anonymity, temporal delay, and social distance—allowing reciprocal logic to function at a societal scale.

\subsubsection{Historical Illustration and Support}

Archaeological records from ancient Mesopotamia suggest that early institutions formalized reciprocal obligations. Temples collected grain as standardized contributions and redistributed it for protection, rituals, or famine relief \cite{graeber2014debt, scott2017against}. Early monetary units recorded on cuneiform tablets served as abstract measures of debt, not as market currency \cite{graeber2014debt}. 

These practices, centered on role-based redistribution and symbolic accounting, appear to have provided the foundation for governance—not through contract, but through the externalization of reciprocal expectations into rules and records.

\subsubsection{Simulation Pathways and Implementation}
Institutions are implemented as fixed, environment-level constraints that agents must comply with during local interactions. Unlike dynamic norms, institutional rules are externalized standards that do not evolve from agent behavior, but instead impose stable structural expectations on the system. The implementation can be specified as follows:

\begin{itemize}
    \item \textbf{Institutional constraint:} A set of fixed, externally defined rules or roles embedded in the environment, which agents perceive and must adapt their behavior to. These constraints are stable across episodes and uniformly accessible to all agents.
\end{itemize}

These constraints structure agent behavior by imposing fixed, shared rules—ensuring consistency across interactions regardless of context or partner.

\section{Implications and Discussion}
\label{others}

\subsection{Limitations and Scope of the Framework}

\paragraph{Empirical Gaps and Historical Limits.}
Our framework draws on ethnographic and historical observations, but such records are often fragmentary, indirect, and temporally sparse—offering limited resolution into the fine-grained mechanisms or sequencing of reciprocal dynamics. Simulation may help fill these empirical gaps.

\paragraph{Toward Broader Institutional Complexity.}
Our framework focuses on the minimal behavioral substrate required for the emergence of scalable social structures. While modern systems—such as legal codes, bureaucracies, or monetary institutions—may involve additional layers of symbolic mediation, cultural learning, or leadership dynamics, these are not addressed here and may be explored in future work.

\subsection{Theoretical Implications}

\subsubsection{Reciprocity as the Scalable Engine of Social Structure}

Most mechanisms that stabilize behavior—such as aversion to norm violation, punitive responses, or in-group bias—function as internal constraints. While they anchor social expectations within familiar contexts, they do not provide a generalizable substrate for large-scale coordination.

Reciprocity, by contrast, enables the circulation of goods, obligations, and cooperative behavior across time and social distance. It allows decentralized societies to function without centralized enforcement—supporting scalable coordination through locally grounded exchanges. From this perspective, institutions are not departures from reciprocity, but its formalized extensions.

\subsubsection{Why Pure Social Reciprocity Fails to Scale to Society}

Anthropological literature often idealizes reciprocity as communal, moral, and stable. However, informal reciprocal systems rely heavily on dense familiarity and continuous exchange. As scale increases, these systems become fragile, coercive, and exclusionary.

What are often seen as moral alternatives to formal institutions may, in practice, create burdens of expectation and reputational pressure. Rather than viewing markets, rules, and norms as deviations from pure reciprocity, we interpret them as structural responses to its limits—mechanisms that externalize expectations, reduce ambiguity, and stabilize cooperation under less personal conditions.

\subsubsection{Power as a Group-Level Force in Structuring Reciprocity}

While reciprocity provides a flexible basis for cooperation, real-world interactions are rarely equal. Over time, differences in influence, status, or resource access begin to shape the dynamics of exchange. Power alters who gives first, who expects more, and how obligations are distributed—embedding durable asymmetries into otherwise mutual interactions.

Power is often treated as symbolic authority or institutional dominance. Yet behavioral research suggests it is rooted in biologically grounded motivational tendencies—such as role assertion, resource control, or status pursuit—that accumulates into stable differentiation over time \cite{anderson2015desire, keltner2003power, thomsen2011big}.

Over repeated interactions, power-linked asymmetries help stabilize and differentiate patterns of reciprocity—turning flexible exchanges into recognizable roles, persistent expectations, and transferrable obligations. This convergence enables local interactions to generalize and scale, laying the groundwork for norm formation and institutional structure.

\subsection{Future Work}

\subsubsection{Operationalizing Social Emergence Stages}

Future work will focus on operationalizing and simulating each stage of social structure emergence from decentralized reciprocal interactions.

\paragraph{Stage 1: Reciprocal Dynamics}
Validate the emergence of stable reciprocal patterns through minimal partner-specific memory and cost–return tracking:
\begin{itemize}
    \item \textbf{Partner-specific memory:} Structured records of interaction outcomes.
    \item \textbf{Reciprocal evaluation:} Basic scoring based on cooperation versus defection.
    \item \textbf{Behavioral updating:} Strategy adjustment based on perceived reciprocity balance.
\end{itemize}

\paragraph{Stage 2: Norm Stabilization}
Examine how decentralized behaviors aggregate into population-level soft norms:
\begin{itemize}
    \item \textbf{Normative macrostate:} Emergent behavioral regularities that guide strategy updates, shape social expectations, and adapt through feedback over repeated interactions.
\end{itemize}

\paragraph{Stage 3: Institutional Construction}
Model the externalization of stabilized norms into fixed environmental constraints:
\begin{itemize}
    \item \textbf{Institutional constraint:} Stable, globally accessible rules embedded in the environment.
\end{itemize}

Together, these pathways provide a mechanistic roadmap for modeling the decentralized emergence of scalable social organization.

\subsubsection{Operationalizing Power as an Internal Motivational Variable}

Power dynamics are widely observed across species and social systems—not as externally imposed structures, but as patterned behaviors shaped by internal motivational tendencies such as status pursuit, resource control, social predictability, and prestige signaling. 
These tendencies influence how individuals navigate interaction, seek influence, and respond to social context, though they may not always manifest directly within reciprocal exchanges.

In our current framework, power is treated as an underlying internal condition—affecting behavior but not yet explicitly represented as a computational variable. 
Future work will focus on operationalizing these motivational substrates within agent-based models, allowing power to influence partner selection, expectation setting, and asymmetry in cooperation. 
This would allow the simulation of how influence and role asymmetries emerge from decentralized settings, advancing a more comprehensive account of scalable social organization.

\section{Conclusion}

Ethnographic and archaeological evidence suggests that reciprocity was the core organizing logic of early human societies—sustaining exchange, cohesion, and obligation without formal institutions. In modern contexts, financial systems such as credit and currency can likewise be interpreted as scalable extensions of reciprocity—enabling cooperation across time, space, and anonymity. 

Drawing on insights from anthropology, history, and behavioral science, we formalize this continuity through a three-stage framework—Reciprocal Dynamics, Norm Stabilization, and Institutional Construction—which reconstructs how complex social systems can emerge from decentralized reciprocal interaction alone, without presupposing centralized control or symbolic consensus.

This framework offers a simulateable, cognitively minimal foundation for understanding the behavioral origins of institutional order. Future extensions may incorporate motivational asymmetries, power dynamics, and learning mechanisms—enriching this foundation toward a more complete model of scalable social coordination.

\section*{Declaration of LLM Usage}
The authors used OpenAI's ChatGPT to assist in refining phrasing and improving clarity. All theoretical arguments and interpretations are original and authored by the researchers.






\bibliographystyle{abbrvnat}
\bibliography{references}

\end{document}